# Possible Outlet of Test on Violation of Pauli Exclusion Principle

*Yi-Fang Chang*
*Department of Physics, Yunnan University, Kunming 650091, China*
(E-mail: yifangchang1030@hotmail.com)

**Abstract**  The present experimental tests have proved high precisely the validity of the Pauli exclusion principle (PEP) in usual cases. The future experiments should be combined widely with various theories of hidden and obvious violation of PEP. Author think that known experiments and theories seem to imply the violation at high energy. Some possible tests have been proposed in particle physics, nuclei at high energy and astrophysics, etc., in particular, the excited high-n atoms and multi-quark states. Moreover, the violation is possibly relevant to the nonlinear quantum theory, in which the present linear superposition principle may not hold. Finally, the possible violation at very low temperature is discussed. These experiments may be connected with tests of other basic principles, for example, the present wave property, possible decrease of entropy due to internal interactions in isolated systems, and PCT invariance, etc.

**1. Small Violation Theory of Pauli Principle and Known Experimental Tests**

In 1987, Ignatiev and Kuzmin constructed a model of a single oscillator violated the Pauli exclusion principle(PEP) possibly. Then, Greenberg and Mohapatra [1,2] generalized to a local quantum field theory, where PEP should have a small violation. Because the violation of PEP is only exactitude, so any natural substance should contain a fraction of order $\beta^2$ (the violation parameter) of anomalous atoms and nucleons, etc. They pointed out: "no high-precision tests of the Pauli principle have been made"[1]. Therefore, many very sensitive experiments and theories are stimulated by this scheme.

But, the after experiments show the validity of PEP with high precision for usual cases. Rahal and Campa [3] discussed a thermodynamics allowing the introduction of a small violation of PEP, and at low temperatures obtained an experimental value $\beta^2 < 10^{-44}$ for X-ray transitions in the electron shells. Gavrin, et al. [4], estimated $\beta \leq 10^{-30}$ for the similar case. Ramberg and Snow [5] searched for anomalous X-rays arising from a small violation of PEP in current carrying copper, but no such signal was found. They derived the limit $\beta^2 < 3.4 \times 10^{-26}$ on possible violation. Drake [6] predicted energy shifts for helium of violation, and an upper limit on the is $\beta^2 \leq 2 \times 10^{-7}$. Plaga [7] inferred a limit on the violation of PEP for the hydrogen burning rate of the solar interion, and proposed that a very small violation in a system of two nucleons might solve the solar neutrino problem. Novikov, et al., [8] used anomalous atoms to test the validity of PEP for atomic electrons, and obtained two limits

$$^{20}\tilde{N}e/^{20}Ne < 2\times 10^{-21}, \,^{36}\tilde{A}r/^{36}Ar < 4\times 10^{-17}. \qquad (1)$$

Arnold, et al., [9] test PEP using the NEMO-2 detector. Limits on the violation of PEP for nucleons and quarks are tested [10,11], and the limits are the order of $10^{-13} - 10^{-18}$ by protons and of $10^{-20} - 10^{-25}$ by neutrons [10].

Moreover, in the theoretical aspect Biedenharn, et al., [12] pointed out that the symmetry underlying PEP could not be violated by an arbitrarily small amount. Govorkov [13] proved that "the Pauli principle is a consequence of only general assumptions of the quantum field theory and small violations of it is not admissible", and "the parameter $\beta^2$ has a fixed finite value and cannot be made arbitrarily small", so "the IKGM scheme cannot be the theory of small violation of the Pauli principle", "The violation of the Pauli principle would mean the inevitable impossibility of it description within the local quantum field theory". According to the IKGM theory, not only the experiments must be very high-precision, but also some mathematical conclusions of quantum theory, whose formulation has self-consistency and PEP is namely its exact result, should be only approximations in usual cases. Then, Greenberg and Mohapatra have found that their theory has difficulties as a local quantum field theory, so it is probably a phenomenological theory introduced a violation parameter $\beta^2$ of PEP.

Now the tests seem to be "no-go" way. But, combining various hidden or obvious violation theories or their mixture, some possible tests under certain particular conditions, for example, at high energy, etc., will be able to be extricated from straits of experiments, perhaps.

## 2. Various Theories on Violation of PEP

Actually, the theories relevant to possible violation of PEP are never only the IKGM scheme [14]. In 1978, Santilli [15] was the first to propose the test of PEP, and then Ktorides, Myung and Santilli [16] pointed out the possible inapplicability of PEP under strong interactions, and possible deviations from PEP can at most be very small. In 1984, based on some experiments and theories of particles at high energy, I suggested that particles at high energy would possess a new statistics unifying Bose-Einstein(BE) and Fermi-Dirac(FD) statistics, and PEP would not hold at high energy [17]. The multiplicity and the large transverse momentum are independent of energy and the types of particles, no matter whether bosons or fermions, corresponding statistics is the $\Gamma$ distribution

$$y = \frac{\beta^\alpha}{\Gamma(\alpha)} x^{\alpha-1} \exp(-\beta x) . \qquad (2)$$

It is unified for BE and FD statistics, and agrees quantitatively with KNO scaling and Dao scaling. The formula (2) can be obtained from the urbaryon parton model or the information theory or the geometric bremsstrahlung model, etc. The scattering cross sections of different particles at high energy tend towards unification. Further, some possible tests of the violation of PEP have been proposed [18,19].

Moreover, the parastatistics is generalized BE and FD statistics; the (2+1)-dimensional nonlinear sigma model solitons may have arbitrary fractional or even irrational spin; the fractional statistics interpolates continuously between BE and FD statistics [20], corresponding anyons interpolate between bosons and fermions, etc. Kivelson [21] discussed these experiments in the fractional quantum Hall state. They have exhibited some contradictions with the standard theory in which two types of different particles and their properties are distinguished from the spin-statistics stringently. Even in the nonabelian gauge field theory there is the ghost particle whose spin is zero, but which agrees with anticommutation relation. Various supersymmetric theories, including well-known superstring, possess a basic symmetry between bosons and fermions, so their formulations are usually unification for both particles. Therefore, these theories correlate possibly to hidden violation of PEP, in which some principles and conclusions of the present quantum theory should be corrected and developed under certain conditions.

## 3. Possible Tests on Violation of PEP

Now all experimental tests are run at low energy region. If future experiments are not confined to high precision, and can widen outlook, and are combined with various theories on violation of PEP, so the violation may not be small or field is nonlocal. Possibly, new tests should perform under some extreme conditions, e.g., at high energy, etc. First, some known experiments at high energy have implied this possibility of the violation of PEP [18]. Next, the above tests of PEP are nonresistant to possible violation at high energy. Thirdly, the asymptotic freedom as a result of many experiments and of the nonabelian gauge field theory seems to have shown that the Pauli exclusion force does not exist among fermions for small distance and corresponding high energy.

I proposed various possible tests of PEP in the following ways [18,19]: the ultrahigh excited state of atoms or nuclei; various nuclei at high temperature, high pressure, high density and at high energy; dineutrons in extremely neutron-rich nuclei; the multiple production at high energy; the internal structure of particles; the gamma-ray sources in high energy astrophysics; the early stage of universe evolution; the black hole and the neutron stars; etc.

The most notable and realizable test is in the excited high-n atoms. For atomic electrons, if PEP is violated, the K shell will be able to accommodate more than two electrons. Rinneberg, et al., obtained high-n Rydberg atoms with the principal quantum number n=290 for in the laboratory [22]. Then they obtained again atoms with n=520. Ling, et al., observed Rydberg state with n=1000 [23,24]. In last case, its high energy level is $10^6$ times as large as "normal"±at omat low energy, and the effective radius is

$$a_n = n^2 \hbar^2 / \mu e^2 = 5.29 \times 10^{-3} cm. \qquad (3)$$

According to quantum mechanics, the electron number in atom must be either two for usual orbit or infinite for ionized state. I believe that there is third possibility: For very high excited atoms, at above near-macroscopic orbit three electrons seems to be able to coexist, at least in a short time interval, which just corresponds to high energy. Moreover, in highly excited atom the effect of spin can be neglected [22], it is just that I expected the condition of the unified statistics and of the inapplicability of PEP at high energy [17]. Further, it is validated that "magic" Rydberg states with n=150 possess enough long lifetimes [25,26].

The present nuclear theory has some difficulties for the nuclear matter at the extreme conditions. The cluster model of the resonating group structure of nucleus has considered sufficiently PEP. If PEP is violated, this model will be different. Now the model is just applied mainly at low energy. Perhaps, the magic number and the shell model at high energy will deviate, BE and FD statistics of nuclei turn towards unification at high energy. Two or more nucleons will be able to fill one state in nuclei at high energy, while the statistical methods and models applied at high energy never exclude a possibility, which allows more than one fermion to possess the same energy. The tests of PEP and the related problems, for example, the PCT symmetry, will be superiority and complexity by nuclei at high energy, since there are numerous various nuclei.

Mohapatra [27] predicted the presence of a neutral spin-3/2 hadron with mass in the 1~2 *GeV* range by using infinite statistics. It implies the violation of PEP at 1~2 *GeV*. I expected that usual high energy is about 2~20 *GeV* for particles according to the uncertainty principle [17]. Greenberg and Mohapatra [1] pointed out that if PEP is violated, neutrons can fall into inner shells, which normally would be filled and gamma rays in the range of 30 *MeV* and above can be emitted. But it is not pointed out that where are the gamma rays of this energy level. Perhaps, the rays may be emitted directly from highly excited nuclei or nuclear collisions at high energy. I discussed possible violation of PEP for the gamma-ray sources, and derived some quantitative results and four conclusions for neutron stars [19].

For the internal structure of hadron at high binding energy the same quarks (e.g., seaquarks), etc., may exist in the same state possibly, so are some subquark models within a quark. If the quark-gluon plasma is produced, the strangelet (multiquark liquid-drop) will be able to use to discuss the violation of PEP. Moreover, various exotic particles and phenomena should be remarked, perhaps, it has implied unifyon, exotion, and parason of the violation of PEP, and they are analogue with (2+1)-dimensional anyon and the ghost particle.

Recently, several groups (LEPS, DIANA, CLAS and BES Collaborations) observed some multi-quark resonances at high energy [28-30]. For example, an exotic baryon $\Theta^+$ (1540) with the quantum numbers of $K^+n$ has been reported, in which five-quark($qqqq\bar{q}$) configurations are mixed with the standard three-quark valence configuration. These multi-quark states coexist inside a short time, which increases a possibility of violation of PEP

## 4. Nonlinear Theory and Violation of PEP

In various researches, the conditions of violation of PEP are different. Santilli's theorie [15,16] predict internal deviations for the constituents of composite systems at mutual distances equal or smaller than 1 fm, while the total spin is conventional. The suppositions [16] are based on that particles under strong interactions need possibly new physical and mathematical generalizations, for instance, Lie-admissible algebra. The IKGM model [1] used the trilinear anticommutation relations

$$a^2 a^+ + \beta^2 a^+ a^2 = \beta^2 a, \qquad (4)$$

etc. I consider that the mathematical formalism of the violation of PEP should be connected with the nonlinear quantum theory [31].

Because of Weinberg's nonlinear quantum mechanics [32], physicists are attaching again importance to the nonlinear problem. Bollinger, et al., [33] discussed a test of the linearity of quantum mechanics, and can set a limit of $4 \times 10^{-27}$ on the fraction of binding energy per nucleon of the Be nucleus that could be due to nonlinear corrections to quantum mechanics. Chupp and Hoare [34] observed coherence among the four magnetic sublevels of freely precessing Ne, and nonlinear corrections to quantum mechanics are found to be less than $1.6 \times 10^{-26}$ of the binding energy per nucleon of Ne. Majumder, et al., [35] obtained that the fraction of nonlinear effects is less than $2.0 \times 10^{-27}$ in optically pumped Hg atoms. Perhaps, the nonlinear effects can be exhibited only for some special cases, for instance, the nonlinear photon-photon interactions in strong field and at high energy.

The present quantum mechanics is based on the superposition principle. "It follows from the principle of superposition of states that all equations satisfied by wave functions must be linear in $\psi$" [36], and the usual quantum theory postulate that all of operators are linear. It is based on the linear superposition principle, or on Fourier transform, or on the association of particles with plane waves. But, we have known that the linear superposition principle and Fourier integral had not held for the nonlinear wave. When particles correspond to nonlinear waves (e.g., solitons) in some cases, the theories will not be linear. Furthermore, in the nonlinear quantum theory the equations and operators are nonlinear [19,31], so the present applied linear superposition principle should be developed, for example, it may be the Backlund transformations of solitons. The quantum-entangled state is namely a nonlinear superposed state. Combining some known results, I proposed a fundamental operator [19,37],

$$p_\mu = -i\hbar (F \frac{\partial}{\partial x_\mu} + i\Gamma_\mu), \qquad (5)$$

where $F$ and $\Gamma_\mu$ are corrected factor and additive term, respectively, and both may be nonlinear forms. From this operator we may derive the commutation relations (quantized conditions)

$$x_\mu p_\nu - p_\nu x_\mu = i\hbar F \delta_{\mu\nu} - i\hbar(x_\mu F \frac{\partial}{\partial x_\nu} - F \frac{\partial}{\partial x_\nu} x_\mu) + \hbar(x_\mu \Gamma_\nu - \Gamma_\nu x_\mu), \quad (6)$$

and the anticommutation relations

$$x_\mu p_\nu + p_\nu x_\mu = i\hbar F \delta_{\mu\nu} - i\hbar(x_\mu F \frac{\partial}{\partial x_\nu} + F \frac{\partial}{\partial x_\nu} x_\mu) + \hbar(x_\mu \Gamma_\nu + \Gamma_\nu x_\mu). \quad (7)$$

If the definitions of corresponding annihilation and creation operators are invariant, Eq.(7) will become $\{a_\alpha, a_\alpha^+\} = F + A$. The number operator is $N_\alpha = a_\alpha^+ a_\alpha$, then

$$N_\alpha^2 = a_\alpha^+ a_\alpha a_\alpha^+ a_\alpha = a_\alpha^+ (F + A - a_\alpha^+ a_\alpha) a_\alpha = a_\alpha^+ a_\alpha (F + A) = N_\alpha (F + A), \quad (8)$$

so $N_\alpha = 0$ and F+A. When F=1 and A=0 (or A<<1), the eigenvalues are 0 and 1. It is low energy cases, and obeys PEP. If the above conditions do not hold, PEP may not hold [19,37].

Otherwise, the anyon is connected with the nonlinear sigma model. The ghost particle corresponds to the nonabelian gauge field, whose equations have nonlinear terms. The equations of the supersymmetric theory mainly are nonlinear. The IKGM theory is combined with the anyon model, probably, there is $\beta^2 = [\exp(-iq\phi/2\pi)]^n$.

**5. Discussion**

At very low temperature two fermions can constitute a boson like the Cooper pair, and perform Bose-Einstein condensation. In 1995, the condensation numbers of $^{87}Rb$ and $^{7}Li$ atoms may be high as $10^5$ under this extreme condition [38]. In this case PEP has not plied a role in the ultracold structure, and violation of PEP may be tested.

Based on the analysis of the logical structure, I think [37], the duality, the wave-property is the basic principle of quantum mechanics. But, in the following relevant four aspects: 1.For single particle, the probability wave has no longer meaning. 2.In an exceedingly small time-space, wave concept itself (for example, wave-length and frequency) does not hold. 3.For strong and weak interactions of short-range. 4.For high energy process, whether the wave property still holds or not, it seems have not been well tested. Under some conditions, the present wave property cannot be exhibited probably, wave should be corrected and developed, and for example, it is nonlinear wave and does not obey linear superposition principle. These are consistent with the possible inapplicability of PEP under strong interactions [16], and with the linearity of quantum mechanics [32-35]. The quantitative restrict of wave property is mainly determined by $\lambda = h/p$. For particles with high energy or large mass, the wavelength is particular small. The relativity is a range of high velocity and large momentum, while microphysics wave property exhibits easily only with low velocity and small momentum. The above-mentioned ranges are contradictory to each other.

So far, any real quark is not found although do the best one can. Therefore, I combined usual cases: various statistical models appear at higher energy, and various symmetrical models appear at lower energy, then suggested that the basic characteristic of particles is the symmetry-statistics duality [37]. So quarks are possibly magic, and are some substeady bound states within hadrons possessed SU(N) symmetry.

In the Universe, time is irreversible, matter and antimatter are asymmetric, the anisotropy has been exhibited, so the parity is not invariant, therefore, the PCT invariance does not hold probably. I proved possible decrease of entropy in isolated systems, whose conditions are the existence of internal nonlinear interactions [39,40].

In short, it will be powerful for the experimental test that the possible conditions and formulations of violation of PEP are discussed widely. In addition, these tests for violation of PEP will be able to apply to test some basic principles and theories.